\documentclass[12pt]{article}

\usepackage{arxiv}
\pdfoutput=1

\usepackage{graphicx}

\usepackage{doi}

\usepackage[toc,page]{appendix}

\usepackage{hyperref}

\newtheorem{theorem}{Theorem}
\newtheorem{property}{Property}

\newenvironment{proof}{\noindent{\bf Proof:} \hspace*{1em}}{
    \hspace*{\fill} $\Box$ }

\newcommand{\handout}[5]{
   \renewcommand{\thepage}{#1-\arabic{page}}
   \noindent
   \begin{center}
   \framebox{
      \vbox{
    \hbox to 5.78in { {\bf ORIE 6334 Bridging Continuous and Discrete Optimization} \hfill #2 }
       \vspace{4mm}
       \hbox to 5.78in { {\Large \hfill #5  \hfill} }
       \vspace{2mm}
       \hbox to 5.78in { {\it #3 \hfill #4} }
      }
   }
   \end{center}
   \vspace*{4mm}
}

\usepackage{amsmath}
\usepackage{amsfonts}
\usepackage{amssymb}

\title{Scheduling Appointments Online:\\ The Power of Deferred Decision-Making}

\author{ 
{
\hspace{1mm}Devin Smedira}\thanks{Work done while a Masters of Engineering Student} \\
	Computer Science\\
	Cornell University
	\And
	{
	\hspace{1mm}David Shmoys} \\
	Operations Research \& Information Engineering\\
	Cornell University
}

\date{}

\renewcommand{\headeright}{}
\renewcommand{\undertitle}{}
\renewcommand{\shorttitle}{The Power of Deferred Decision-Making}

\hypersetup{
pdftitle={Scheduling Appointments Online: The Power of Deferred Decision-Making},
pdfauthor={Devin ~Smedira, David ~Shmoys},
}

\begin{document}
\pagenumbering{roman}
\maketitle

\begin{abstract}
The recently introduced online Minimum Peak Appointment Scheduling (MPAS) problem is a variant of the online bin-packing problem that allows for deferred decision making. Specifically, it allows for the problem to be split into an online phase where a stream of appointment requests arrive requiring a scheduled time, followed by an offline phase where those appointments are scheduled into rooms. Similar to the bin-packing problem, the aim is to use the minimum number of rooms in the final configuration. This model more accurately captures scheduling appointments than bin packing. For example, a dialysis patient needs to know what time to arrive for an appointment, but does not need to know the assigned station ahead of time.

Previous work developed a randomized algorithm for this problem which achieved an asymptotic competitive ratio of at most 1.5, proving that online MPAS was fundamentally different from the online bin-packing problem. Our main contribution is to develop a new randomized algorithm for the problem that achieves an asymptotic competitive ratio under 1.455, indicating the potential for further progress. This improvement is attained by modifying the process for scheduling appointments to increase the density of the packing in the worst case, along with utilizing the dual of the bin-packing linear programming relaxation to perform the analysis. We also present the first known lower bound of 1.2 on the asymptotic competitive ratio of both deterministic and randomized online MPAS algorithm. These results demonstrate how deferred decision-making can be leveraged to yield improved worst-case performance, a phenomenon which should be investigated in a broader class of settings.
\end{abstract}

\newpage
\pagenumbering{arabic}



\section{Introduction}

The bin-packing problem is a classic and well-studied problem in algorithm design, with applications in a wide array of industries \cite{johnson1974fast, csirik1998line, coffman2013bin}. The problem requires packing items of given lengths into unit-length bins such that no two items in a bin overlap, with the goal of minimizing the total number of bins used. In particular, the ``online'' version of the problem is historically important and continues to be practically relevant. In this version, items are received one at a time, and each item must be placed in a bin before the next item is received. Existing approximation algorithms for it have become very well refined, with upper and lower bound results on the competitive ratio becoming very close in the last few years \cite{balogh2021new, balogh2017new, chandra1992does}. Further, the optimal absolute competitive ratio is known \cite{balogh2019optimal}. 

This paper studies a variant of the well-known bin-packing problem called the minimum peak appointment scheduling (MPAS) problem \cite{escribe}. Recently proposed, an input to this problem is a set of jobs, each of a given length, that needs to be scheduled within a period of time. For example, one might be assigning each job to an interval within one 9AM-5PM work day. The objective is to minimize the peak utilization, that is, the maximum number of jobs being serviced simultaneously. In fact, Escribe et al. \cite{escribe} highlight this objective, since they considered an application in which each job is also assigned to a facility, and the aim is to minimize the number of facilities needed. One can view each facility as a bin, and the position within the bin as the time interval to which a job is assigned. They observe that the minimum number of bins required (for a feasible schedule) is equal to the peak utilization, which follows directly from the fact that interval graphs are perfect (see, e.g., \cite{golumbic2004algorithmic}). Therefore, in the offline setting, this scheduling problem is identical to the bin-packing problem. In the online setting, however, as requests for appointments arrive one after the other, only a time interval needs to be committed to, instead of a facility (as well as the time). This allows for added scheduling flexibility which can be exploited. As Escribe \cite{escribe} showed, this difference is sufficient to prove that the online MPAS problem is fundamentally different from bin packing. This is highlighted in Table 1, where the column denoted asymptotic ratio, referring to the asymptotic competitive ratio, gives a proven upper or lower bound on the ratio between objective function of the solution found to the optimal (taken in the limit over increasing input sizes). 

\small
\begin{center}
Table 1: Comparison of previously known bin-packing and MPAS bounds
 \begin{tabular}{||c c c c c||} 
 \hline
Problem & Type of Algorithm & Type of Bound & Asymptotic Ratio & Paper \\ [0.5ex] 
 \hline\hline
Bin-Packing & Deterministic & Lower Bound & 1.54278 & Balogh et al. (2021) \cite{balogh2021new} \\ 
 \hline
Bin-Packing & Deterministic & Upper Bound & 1.57829 & Balogh et al. (2017) \cite{balogh2017new} \\
 \hline
Bin-Packing & Randomized & Lower Bound & 1.536 & Chandra (1992) \cite{chandra1992does} \\ 
 \hline
MPAS & Randomized & Upper Bound & 1.5 & Escribe et al. (2021) \cite{escribe} \\ 
 \hline
\end{tabular}\\
\end{center}
\normalsize

Many real-world decision-making problems can be solved by framing them as a bin-packing problem and applying existing algorithms. However, using the familiar but rigid framework of bin packing can create solutions that do not fully utilize the flexibility of the particular setting at hand.
For example, an infusion center might need to determine how many chairs to configure each day. As appointments are made, the center may only need to commit to a particular time for the appointment - they do not necessarily have to commit to a particular chair for that appointment. An event host might need to determine how many rooms are necessary to host a particular event, but they may not need to tell the attendees which room they will be in until the day of the event. In both cases, traditional bin-packing algorithms can provide answers to these questions, but as this work will show, these solutions may result in an unnecessary waste of resources. In this sense, the MPAS framework is a natural generalization of the traditional bin-packing paradigm that is tailored for optimization problems that revolve around appointment scheduling. While this framework can be used to improve the solutions generated by approximation algorithms, the principles developed by the exploration of this framework can also be used to improve the efficiency of operations even when the scheduling is managed manually.

In additional to its practical use, this problem is also of theoretical interest because it is an effective vehicle for studying the power of delayed decision-making in online problems. This problem is partitioned into two fundamental steps, the first conducted in real time online and the second conducted later offline. One goal of this work is to highlight how this partition can be leveraged to improve on the solution to the normal online bin-packing problem, with the hope that similar delayed decision principles might lead to improvements in the solutions to other existing or future online problems. In particular, one could imagine other online bin-packing relaxations arising in different settings, the study of which would be aided by work on the online MPAS problem.

To underscore the connection of this work to bin-packing problems, this paper will use terminology commonly found in the bin-packing literature. Arriving jobs for the MPAS problem will be referred to as items, and their duration referred to as their size. Scheduling an item will be referred to as placing it into a position within a bin. While it is traditional to represent bin-packing solutions in which the bins are oriented vertically, in this paper, we will be view them as oriented horizontally, with position 0 on the left and position 1 on the right; this orientation provides a more intuitive connection to the time intervals the bins represent, analogous to a Gantt chart representation for a schedule. For example, scheduling an item at the left end of a bin is analogous to the first appointment of the interval, and scheduling an item to the right of another item corresponds to the item occurring later.

\paragraph{An Overview of Known Algorithms}

In addition to defining the MPAS problem, Escribe et al. \cite{escribe} contributed a novel randomized algorithm for the online MPAS problem called the Harmonic Rematching Algorithm and provided an analysis to prove the algorithm had an asymptotic competitive ratio of 1.5. 
The algorithm that we propose and analyze in this work is greatly inspired by the Harmonic Rematching algorithm, which itself was influenced by the harmonic algorithm for online bin-packing \cite{lee1985simple}. Like the problem, the algorithm is composed of two phases: an initial scheduling phase done in an online fashion and a subsequent rematching phase that places items into final bins. The number of bins used in the algorithm provides an upper bound on the peak appointments, from which an asymptotic competitive ratio for their algorithm can be calculated.

In the scheduling phase of the algorithm of \cite{escribe}, items are initially sorted into categories based on their size of the form $(\frac{1}{n+1}, \frac{1}{n}]$. Items are then scheduled according to unique rules for each category. Then, during the rematching phase of the algorithm, items of size under 0.5 are rematched into bins containing only other items of the same category and items of size over 0.5. The analysis of the algorithm then shows that an average bin density of $\frac{2}{3}$ is maintained so long as there are under 2 items of size over 0.5 for every 3 bins the algorithm outputs. The derivation of the asymptotic competitive ratio follows directly from this property.

\paragraph{New Contributions}

The primary contribution of this work is to refute a conjecture \cite{Levi} that no algorithm could achieve a lower asymptotic competitive ratio than 1.5 for the online MPAS problem. The new bound is achieved through modifications to the existing algorithm to increase the packing density and the use of a new analytic framework inspired by a primal-dual analysis of the bin-packing linear program. By refuting this conjecture, and providing a more sophisticated algorithm and analysis to do so, this work represents a step towards better understanding the degree to which deferred decision making can improve the performance of online algorithms more broadly.

A secondary contribution of this work was to present the first known lower bounds on the asymptotic competitive ratio of any algorithm for the online MPAS problem. Using a family of inputs consisting only of items with size $\frac{1}{3}$ and $\frac{2}{3}$, it is possible to prove no deterministic or randomized algorithm for the online MPAS problem can achieve an asymptotic competitive ratio below 1.2. For deterministic algorithms, this is relatively straightforward - an adversary can present $n$ items of size $\frac{1}{3}$ first, and depending on whether the algorithm leaves sufficiently many bins with only 1 of these, decides whether to add to the input $n$ items of size $\frac{2}{3}$. The lower bound for randomized algorithms extends this approach using Yao's framework for proving randomized lower bounds \cite{borodin2005online}.

The majority of this paper will be dedicated to outlining the new algorithm for the online MPAS problem and providing an analysis to prove it achieves an asymptotic competitive ratio of $\frac{16}{11} \approx 1.455$. In particular, Section 2 will detail the behavior of the algorithm, whereas Section 3 will present an analysis of the desired performance guarantee. Section 4 will present the lower bound results, whereas Section 5 outlines the potential for further work.

\section{Packing Algorithm}

The following is a detailed description of the algorithm for solving the online MPAS problem. The algorithm will proceed in two main steps. First, there will be a section dedicated to scheduling the items as they arrive. Then, there will be a second separate section dedicated to rematching the scheduled items into final bins. 

An online algorithm for the MPAS problem would be complete with only the first schedule creation phase of the algorithm, since the task is only to schedule appointments to minimize the peak. However, many practical applications would require some mechanism to take this appointments and actually group them together, say to schedule them to a room. For this reason the work includes the second algorithmic step, rematching. This step will also play a role in the analysis of the algorithm, with the number of bins produced by the rematching portion used in place of the number of appointments at the peak time, since the former is an upper bound on the latter.

\subsection{Schedule Creation}

The first phase of this online algorithm is the scheduling phase. During the scheduling phase, new items arrive in an online manner and each item is assigned to a position within a bin.

\paragraph{Item Categories}
Every item will be given a type based on its size. How an item is initially scheduled, as well as in what manner it is rematched into bins, will depend on which type of item it is. These different item types are:

\textbf{Small Item} -  Item of size $\leq 1/3$\\
\textbf{Third Item} - Item of size in $(1/3, 0.34375]$\\
\textbf{Medium Item} - Item of size in $(0.34375, 0.5]$\\
\textbf{Large Item} - Item of size in $(1/2, 0.6875]$\\
\textbf{Very Large Item} - Item of size $\geq 0.6875$

Within the specified range of each item type, there will be one or more different categories. Each category will have its own specific procedure on how items from the category should be both initially scheduled and eventually rematched into bins. The full list of item categories appears in Appendix C, along with other relevant details. Two categories in particular will be important enough in the analysis to separately name. These categories are: 

\textbf{Quarter Items} - A Small Item category for items of size in $(1/4, 0.27]$\\
\textbf{Half Items} - A Medium Item category for items of size in $(0.46, 1/2]$ 

In particular, Quarter Items are important because they cannot be densely packed, and Half Items will be important because 1 Half Item cannot fit in a bin with two Quarter Items.

\paragraph{One-Sided Bins}
The initial scheduling phase of the algorithm will consist of creating a set of \textbf{One-Sided Bins} that will later be rematched. A one-sided bin is a category specific construct containing only items from one particular category. Items will be scheduled into these one-sided bins, and every item in the same one-sided bin will be scheduled in the same bin at the end of the algorithm (with a slight exception for Third Items detailed later). Specifically, each final bin returned by the algorithm will consist exclusively of items from one or two one-sided bins. 

There will be three types of one-sided bins used by the algorithm. They are:

\textbf{Type 1 Bins} - A type 1 bin will be a one-sided bin that is over 0.6875 full. These bins will not be matched with other one-sided bins, so each type 1 bin will be assigned its own bin when the algorithm terminates.

\textbf{Type 2 Large Bins} - A type 2 large bin will be a one-sided bin which is over 0.3125 full but under 0.6875 full. These bins might or might not be matched during the course of the algorithm and so might be paired with another one-sided bin in the algorithm's output.

\textbf{Type 2 Small Bins} - A type 2 small bin will be a one-sided bin that is at most 0.3125 full. All but a constant number of these bins will be matched with another Type 2 one-sided bin, and share an output bin when the algorithm terminates.

Each Small Item category will have a specified number of items per type 1, type 2 large, and type 2 small bin listed in the appendix. Very Large Items will be packed exclusively into type 1 bins containing 1 item, while Large and Medium Items are packed exclusively into type 2 large bins containing 1 item. Third Items will be exclusively packed into type 2 large bins, although some will contain 1 item while others will contain 2. Further, the type 2 bins containing 2 items may be shuffled so that the items initially in the same one-sided bin are not packed in the same bin when the algorithm terminates. However, every Third Item will end in a bin with the same number of items as the one it was initially assigned to. The exact packing mechanism is detailed below.

Type 2 bins will have a further designation of being a left or right sided bin. Simply, a left sided type 2 bin will have the first item placed at position 0 with subsequent items placed closer to 1, while a right sided type 2 bin will have the first item placed at position 1 with subsequent items placed closer to 0. The scheduling process detailed below will ensure some amount of balance between the left and right sided bins.

\paragraph{Complete Sets of One-Sided Bins}
Items from each particular category will be packed into \textbf{Complete Sets of One-Sided Bins}, referred to interchangeably as complete sets. For each item category, a complete set of one-sided bins will be a certain number of type 1, type 2 large, and type 2 small bins. Further, every complete set will contain an even number of type 2 small and type 2 large bins, with half packed on either side. For Very Large Items, a complete set will have 1 type 1 bin, while for Large and Medium Item categories a complete set will have 2 type 2 large bins.  For Quarter Items, a complete set will have 1 type 1 bin containing 3 items, 2 type 2 large bins containing two items, and 4 type 2 small bins containing 1 item.

For Third Items, a complete set will have 4 type 2 large bins containing 1 item each, and 2 type 2 large bins containing 2 items each. These will be referred to as \textbf{Outer Bins} and \textbf{Inner Bins} respectively. This distinction will be important in later analysis. Further, unlike other one-sided bins with multiple items, it will be important for inner bins to be packed so that either item can be placed first. Therefore, the interior item of inner bins will be place exactly 0.34375 from the left or right border, depending on the side of the bin, to accommodate another Third Item of any size. This will also allow any interior left item in an inner bin to be paired with any exterior left item in an inner bin of the same side.

During the scheduling phase, items arrive one at a time. Therefore, it is not always possible to maintain complete sets throughout the algorithm's execution. A \textbf{Partial Set of One-Sided Bins} will refer to a complete set of one-sided bins that has been allocated by the algorithm but has not yet had every item spot filled.

\paragraph{Arrival Scheduling}
This section will detail the actual procedure used for the scheduling phase of the algorithm. When an item arrives, the first step to scheduling is to determine of what category it is a member. Then, the algorithm determines whether or not a partial set of matched bins exists for that category, creating a new one if one does not exist. Then,

For \textbf{Very Large} items, a partial complete set only contains 1 type 1 bin, so the item is placed in it and the set is marked as being completed.

For \textbf{Large and Medium} items, a complete set of bins only contains two 1 item bins. So if the partial set has one of the type 2 large bins filled, the other bin is filled with the item and the set is marked as completed. Otherwise, the side of the bin the item is placed on is chosen randomly with equal probability.

For \textbf{Third} items, the item scheduling process is several steps. Within the partial set, if there are more items in inner bins than outer bins then mark the new item for an outer pair and vice versa. If the item is marked for an outer bin, place it in a type 2 large bin on the side with fewer items, or randomly with equal probability if there are the same number of items on each side. If the item is marked for for an inner bin, first choose whether to place it in the left or right side bin based on which has fewer items, or randomly if they are equally full. Finally, place it on the edge of the bin if there is an item in the interior already and vice versa. Again choose randomly if both spots are not filled. If there are no more open spots in the partial set, mark the set as completed. This complicated procedure will play an important role in future analysis.

For \textbf{Small} items, items can be placed arbitrarily among the open spots within a partial set of the correct category. If there are no more open spots in the partial set after the item is placed, mark the set as completed.

The above procedure will ensure that each item category has at most 1 partial set of one-sided bins at any time during execution. Thus, when the scheduling phase terminates, there will be a finite number of partial sets, allowing them to be ignored in the final analysis.

\subsection{Rematching}

The second phase of the algorithm will be an offline rematching phase. This step will take all of the items that are already committed to a position within the bin and assign them to final bins in a manner such that no two items within a bin overlap. For the sake of this overview and the analysis, it will be sufficient to simply say that items from some category were rematched with items from another category. The exact details of how the rematching puts items into bins is detailed in appendices B and C.

\paragraph{Large and Very Large Items}
Very Large Items are exclusively put into type 1 bins, and as such will not need to be rematched by the algorithm. Each Very Large Item will end up in its own bin when the algorithm finishes.

Large Items are exclusively put into type 2 large bins, and can not fit in a bin with another Large Item. They will be used to rematch with smaller items, detailed below. Though complete sets of Large Items contain two Large Items, the individual items may be rematched with different sets during the rematching process.

\paragraph{Medium Items}
As is the case with Large Items, Medium Items may not be packed with the other item originally in the same complete set. Rather, each item will be used individually in the rematching stage. A Medium Item can be matched in a bin with another Medium Item of the same category or matched in a bin with a Large Item on the opposite side if the Large Item is small enough to fit with the Medium Item. At most 1 Medium Item per category will not be matched in one of these two ways. Depending on the distribution of items, five bins of Medium Item pairs may be grouped with 11 bins containing single Large Items for the algorithm's analysis. 

Medium Items not in the Half Item category can also be rematched with Large Items and a complete set of Quarter Items.

\paragraph{Third Items}
As is the case with Large and Medium Items, Third Items in the same complete set initially may not be placed in the same group of bins during the rematching process. In fact, Third Items in inner bins may end up placed in an inner bin with a different item than originally matched with. However, Third Items will always be matched in units of complete sets, unlike the Large Items which may not be.

Third Items can be rematched with Large Items if the Third Items in the outer bins and the Large Items can fit in a bin. In this case, the four outer bins will be matched with a Large Item in a bin, the two inner bins will be assigned their own bin, and a 5th Large Item will be assigned its own bin and grouped with the rematched set. Third Items may also be grouped with Large Items that do not fit in a bin with the outer bins. In this case, 11 bins containing a Large Item will be grouped for every 5 bins containing two Third Items. Complete sets of Third Items may also be rematched with complete sets of Small Items as detailed below.

Third Items can also be rematched with Large Items too large to share a bin with and a complete set of Quarter Items.

\paragraph{Small Item Categories}
During the rematching process, every complete set of Small Items may or may not be rematched with other complete sets to assign one-sided bins to bins. Any complete set of Small Items not rematched with other items will have every one-sided bin assigned to bins only with other one-sided bins from the same set. This set bins will be referred to as a \textbf{Complete Set of Matched Bins}. 

All Small Item categories can have their complete sets grouped with other complete sets from different categories to form a \textbf{Complete Set of Rematched Bins}. A complete set of rematched bins will consist of bins containing some combinations of one-sided bins from a complete set of Small Items and one-sided bins from a different item category. The terms complete set of rematched bins and a complete set of bins rematched with, followed by whatever the bins are rematched with, will be used interchangeably.

For any small category besides the Quarter Item category, a complete set of bins can be rematched with Large Items, one or more complete sets of Third Items, Large Items and one or more complete sets of Quarter Items, or one or more complete sets of Third Items and one or more complete sets of Quarter Items.

In addition to all the previously listed cases, a complete set of Quarter Items can also be rematched with Large Items.

\paragraph{Rematching Procedure}
The algorithm will use the following outlined procedure. 

\textbf{Step 1} - Rematch Large Items with Third Items\\
Sort both the left and right Third Items in outer bins in increasing order, and do the same for all four item positions in the inner bins. Also sort the left and right Large Items in increasing order.

As long as there remains complete sets of Third Items and Large Items on both sides, check if the first two outer left Third Items fit in a bin pairwise with the first two right Large Items, and vice versa. If they do not, then move to the next step. If they do, place the outer one-sided bins in a final bin with the Large Item. Then form a right and left inner bin by taking the first item from each respective list and place both inner bins in a bin alone. Finally, take the largest Large Item from whichever side has more items (arbitrarily if a tie) and place it in a bin by itself. 

\textbf{Step 2} - Rematch Large Items with Medium Items\\
Sort both the left and right Medium Items in increasing order, and do the same for the left and right Large Items.  As long as there remains Large and Medium Items on both sides, check if the first left Medium Item fits in a bin with the first right Large Item and vice versa. If they do not, move to step 3. Otherwise, place both pairs into a bin together and repeat.

\textbf{Step 3} - Rematch Large Items with Third and Quarter Items\\
Sort both the left and right Third Items in outer bins in increasing order, and do the same for all four item positions in the inner bins. Also sort the left and right Large Items in increasing order. As long as there remains complete sets of Third Items, complete sets of Quarter Items, and enough Large Items on both sides, do the following. Take 7 Large Items from each side, a complete set of Quarter Items, and a complete set of Third Items and rematch them together. 

\textbf{Step 4} - Rematch Large Items with Medium and Quarter Items\\
Sort both the left and right Medium Items in increasing order, and do the same for the left and right Large Items. As long as there remains non-Half Item Medium Items on each side, complete sets of Quarter Items, and enough Large Items on both sides, do the following. Take 4 Large Items from each side, a complete set of Quarter Items, and one Medium Item from each side and rematch them together. 

\textbf{Step 5} - Rematch Large Items with Complete Sets of Small Items \\
While there remains complete sets of Quarter Items and complete sets of small non-Quarter Items, rematch them together with the appropriate number of Large Items. Do this until there are not enough Large Items or complete quarter sets, or until all non-quarter complete sets have been used.

If complete sets of non-quarter Small Items and enough Large Items remain, continue rematching them until one runs out.

\textbf{Step 6} - Rematch Third Item sets with Complete Small Item Sets\\
While there remains complete sets of Quarter Items and complete sets of small non-Quarter Items, rematch them together with the appropriate number of complete Third Item sets. Do this until there are not enough complete Third Item sets or complete quarter sets, or until all non-quarter complete sets have been used.

If complete sets of non-quarter Small Items  and enough Third Item sets remain, continue rematching them until one runs out.

\textbf{Step 7} Rematch Large Items with Complete Sets of Quarter Items\\
Rematch Large Items with complete sets of Quarter Items until one runs out.

\textbf{Step 8} - Rematch Complete Sets of Third Items with Complete Sets of Quarter Items\\
Rematch complete sets of Third Items with complete sets of Quarter Items until one runs out.

\textbf{Step 9} - Group Large Items with Third Items and Medium Items. \\
While there remains Large Items and either Medium or Third Items that are not placed in a bin, group bins containing two Third/Medium Items with bins containing Large Items at a ratio of 5 third/medium bins to 11 large bins.

\textbf{Step 10} - Assign Remaining Large and Very Large Items to Their Own Bins

\subsection{Example Category Packing}

In the interest of clarifying the algorithm the following section will be dedicated to describing the matching procedure for the category of items in the range $(0.215, 0.23]$. 

\paragraph{Matched Set}
A complete set of items from this category will consist of four type 2 large bins each containing two items and two type 2 small bins each containing 1 item. A complete set of matched bins will be formed from this category by placing both type 2 small one-sided bins in a bin together, and by putting the four type 2 large one sided bins into two bins. Figure 1 below demonstrates how this matching is done, with the blue representing items from the category and the grey representing empty space in the bin.

\begin{figure}[h!]
  \includegraphics[width=\linewidth]{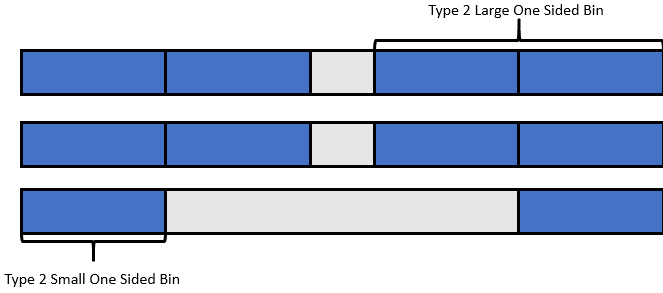}
  \caption{A Complete Set of Matched Bins (Not to Scale)}
  \label{Matched Set}
\end{figure}

\paragraph{Rematched Set}
To fully specify the behavior of rematching with this item category, five different rematching alignments would need to be specified. These would be the complete set rematched with\\
1) Large Items of size under 0.54\\
2) Large Items of size over 0.54\\
3) A complete set of Third Items\\
4) Large Items and a complete set of Quarter Items\\
5) A complete set of Third Items and a complete set of Quarter Items\\
Figure 2 demonstrates case 2 on the left and case 4 on the right, with blue representing items in the category, orange representing Quarter Items, red representing Large Items, and grey representing empty space in the bins. The presented matchings will only have a Large Item in 2 out of every 3 bins, instead of the later purported 0.6875 Large Items per bin. This can be rectified by assigning an additional Large Item for every 15 rematched bins, a process detailed in the appendix. 

\begin{figure}[h!]
  \includegraphics[width=\linewidth]{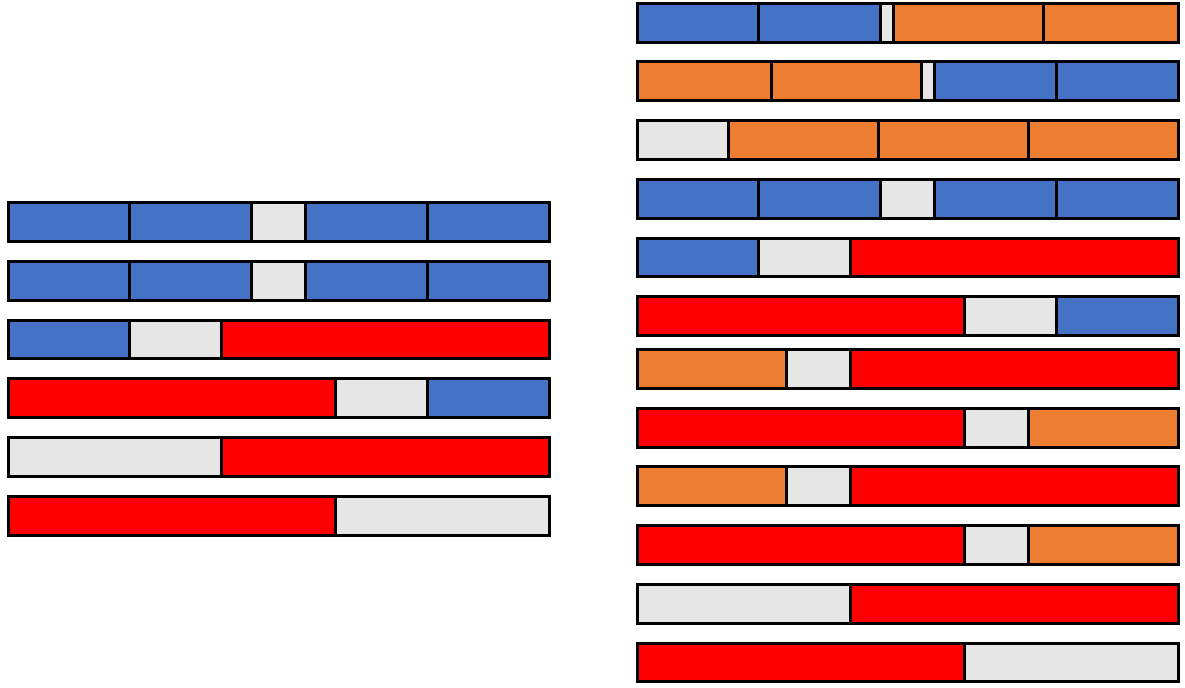}
  \caption{Example Complete Sets of Rematched Bins (Not to Scale)}
  \label{Rematched Set}
\end{figure}
Different item categories will have different packing protocols and different cases than the 5 listed above. But, they will all follow this general pattern on small type 2 bins being matched with Large Items (or 2 Medium Items), while type 2 large bins are matched with small Large Items, each other, or Quarter Item type 2 large bins.

\section{Analysis}

In this section, we will upper bound the asymptotic competitive ratio of our algorithm.
More formally, the asymptotic competitive ratio is defined as 
\[\lim_{N \rightarrow \infty} \sup_{I : OPT(I) \geq N} \frac{ALG(I)}{OPT(I)}\]
where $I$ ranges over all possible inputs to the MPAS problem, $OPT(I)$ is the number of bins used in an optimal solution to the MPAS problem instance $I$ (equivalent to the number of appointments at the peak time), and $ALG(I)$ is the number of bins used by the proposed algorithm for that instance. Notice that the rematching phase of the algorithm is not guaranteed to use the fewest number of bins possible for the given schedule, so $ALG(I)$ might actually be an overestimate of the MPAS objective of the algorithm's scheduling.

\begin{theorem} [Asymptotic Competitive Ratio]
The asymptotic competitive ratio of the algorithm presented in Section 2 is at most $\frac{16}{11}$.
\end{theorem}

To prove the bound, the analysis that follows will show that 
\[\forall I, \frac{11}{16} ALG(I) \leq OPT(I) + K(I)\]
for some function $K(I)$ satisfying $\lim_{OPT(I) \rightarrow \infty} \frac{K(I)}{OPT(I)} = 0$, implying the desired asymptotic competitive ratio. 

The analysis that follows is based on the dual of a bin-packing linear programming relaxation (often referred to as the configuration LP for bin-packing). For each of five separate possible scenarios in which the rematching process might end, a feasible solution to the dual program will be created such that the dual objective $D$ will satisfy $\frac{11}{16} (ALG(I) - K(I)) \leq D$, which will in turn imply that $\frac{11}{16} (ALG(I) - K(I)) \leq OPT(I)$ by weak duality and the fact that the linear relaxation optimal solution is a lower bound $OPT(I)$. This will be sufficient to prove the result.

\subsection{Properties}

In order to conduct the analysis of the algorithm's performance, a few core properties of the algorithm must be established. Each of the following properties have been verified for each of the relevant categories.

\begin{property} [Average Fullness]
For each Small Item category, a complete set of matched bins will have an average fullness of at least 0.6875 per bin.
\end{property}

\begin{property} [Rematched Fullness]
For each Small Item category excluding Quarter Items, a complete set of bins rematched with Large or Third Items will have an average fullness of at least 0.6875 per bin. Further, this packing will have at least 0.6875 Large Items for every bin or 1.375 Third Items for every bin on average. 
\end{property}

\begin{property} [Quarter Rematched Fullness]
For each Small Item category excluding Quarter Items, a complete set of bins rematched with Quarter Items and Large or Third Items will have an average fullness of at least 0.6875. Further, this packing will have at least 0.6875 Large Items for every bin or 1.375 Third Items for every bin on average. 
\end{property}

\begin{property} [Quarter Size Reallocation]
For each Small Item category excluding Quarter Items, a complete set of bins rematched with Quarter Items and Large or Third Items will satisfy $0.3125q + 0.34375t + 0.6875l \geq 0.6875b$, where
\begin{itemize}
\item $q$ is the number of items of size in $(1/4, 1/3]$;
\item $t$ is the number of Third Items in the set;
\item $l$ is the number of Large Items in the set;
\item $b$ is the number of bins the set uses.
\end{itemize}
Complete sets of Quarter Items will satisfy this property whether or not the are rematched with something else.
\end{property}

Like the exact details for matching each item category, the complete verification of these four properties appears in the appendix. These properties will be used in the remainder of the work.

\subsection{Configuration Bin-Packing LP}

The analysis will rely on providing a bound on the minimal number of bins the items could be packed in using a bin-packing linear program relaxation. Define $C$ to be the set of all valid configurations of items from an instance to the MPAS problem, where a {\it configuration} is a set of items that can be feasibly packed together within one bin. Let $I$ be the set of all items in the MPAS instance, with each individual element receiving its own entry; in other words, even if there are two items of the same size, they will be viewed as distinct items. Further define $C(i)$ to be the set of configurations containing the element $i$. One possible linear program relaxation of the bin-packing problem is as follows:
\begin{align*}
\text{Minimize } & \sum\limits_{c \in C} y_c \\
\text{Subject To } & \sum\limits_{c \in C(i)} y_c \cdot c(i) \geq 1 & \forall i \in I\\
& y_c \geq 0 & \forall c \in C
\end{align*}
Intuitively, the $y_c$ variables in this program will represent the number of bins which are packed according to configuration $c$ in a solution. Since this is a relaxation, it is entirely possible that a particular $y_c$ is not an integer value. However, since each configuration contains specific items, no optimal solution will ever have $y_c > 1$. The dual of this program will be:
\begin{align*}
\text{Maximize } & \sum\limits_{i \in I} x_i \\
\text{Subject To } & \sum\limits_{i \in c} x_i \leq 1 & \forall c \in C\\
& x_i \geq 0 & \forall i \in I
\end{align*}
This program will be the one utilized to complete the analysis. Each dual variable $x_i$ intuitively represents assigning a size to a particular item. In this sense, it will be possible to give more or less weight to certain specific items to change the dual objective. The set of additive constraints essentially says that no valid bin configuration can have the weights of its items sum to more than 1. It will be convenient going forward to label the size of an item $i$ as $w_i$.

\subsection{Additive Constants}

Throughout the algorithm's execution, there are several stages where items from a specific category will be used in some quantity. These shortages, detailed below, will not play a role in the final analysis.

\paragraph{Small Category Imbalance of Size-Dependent Scheduling}
Some Small Item categories have analysis of fullness dependent on being rematched with Large Items either all larger or all smaller than a certain threshold. This may not be possible when rematching, but if that is the case then there must be a finite number of Large Items left. Each such threshold is a boundary between different Large Item categories, so this problem cannot arise from an imbalance of item sizes on each side. Thus, if this problem arises, every remaining Large Item can be placed into its own bin and ignored by including those bins in the additive constant for the analysis.

\paragraph{Partial Sets}
Each category can only have at most 1 partial set of one-sided bins at any point during the scheduling phase. Thus there will be only a finite number of items in partial sets when the scheduling phase ends using only a finite number of bins. These bins can be grouped into the additive constant and ignored in the analysis.

\paragraph{Insufficient Items to Rematch}
The rematching phase has several sections with instructions that simply state something along the lines of rematch this item group with some other specified item group. This may not be possible even if items still remain if, for example, there are only 2 Large Items left but 6 needed.  But, if such a problem arises, the category causing the problem will have only finitely many items left, so they can be placed in their own bins and grouped into the additive constant for the analysis.

\subsection{Bounded Imbalance}

In the following sections where the multiplicative bound is derived, the following property will be necessary in some cases of the analysis:

\begin{property} [Balanced Bins]
After the rematching phase of the algorithm is complete, all Large and Very Large Items not matched to a bin with a Medium or Third sized item can not fit in a bin with any Medium sized item not matched to a bin with a Large Item. Further, all Large and Very Large Items not matched to a bin with a Medium or Third sized item can not fit in a bin with any Third sized item in a complete set which does not have a Large and Third Item sharing a bin in its final matching.
\end{property}

It is relatively easy to see that this property does not necessarily hold in the described algorithm. The remainder of this section will be dedicated to proving that 
\[\lim_{N \rightarrow \infty} \sup_{I : OPT(I) \geq N} \frac{IMB(I)}{OPT(I)} = 0\]
where $IMB(I)$ represents the number of items that do not satisfy the balanced bins property. With this proven, the analysis can be completed by assuming the balanced bins property and taking items that defy it and placing them into their own bins. These extra bins will be grouped into the $K(I)$ component of the competitive analysis.

\paragraph{Large and Medium Jobs}
Take some instance of the MPAS problem which has $L_c$ Large Items in each large category $c$ and $M_c$ Medium Items in each medium category $c$. For any large category $c$ and any $x$, define $L_c(x)$ to be the number of large jobs of size less than or equal to $x$ placed on the right side minus the number of such jobs placed on the left in category $c$. Define $M_c(x)$ similarly for medium jobs.  

By Lemma 7 in Escribe\cite{escribe}, there is some constant $\kappa_c$ independent of $x$ and $L_c$ such that with probability 1,
\[|L_c(x)| \leq \kappa_c \sqrt{L_c \log \log L_c} \text{ } \forall x,c\]
along with an analogous result for Medium Items. Further, it must be the case that the optimal solution to this instance is at least $L_c$ (or at least $\frac{M_c}{2}$) for each category $c$. Since $\lim_{L_c \rightarrow \infty} \frac{|L_c(x)|}{L_c} = 0$, the imbalance of Large Items with respect to some cutoff in any category will satisfy the desired property. The same is true for Medium Items.

Now, consider the first time a Large and Medium Item on opposite sides cannot be matched together in the same bin. Assume without loss of generality this happens on the left for the Medium Items and right for the Large Item. At that point, the only items that would not satisfy the balanced bins property would be Large Items on the left and Medium Items on the right in the same category as the items which did not fit together (because the items are guaranteed to be balanced between categories). The number of such items will be precisely the measured imbalance above, so they (and corresponding items on the opposite side to keep the number of items on either side balanced) can be placed in their own bins and ignored for the rest of the analysis.

\paragraph{Third Items}
The proof of the property for Third Items follows the same pattern as the above section, but applies the lemma all four times there is a binary split on item placement. Then, the number of Third Items which do not satisfy the Balanced Bins property can be bounded above by the sum of the 4 imbalances. These imbalances will again approach zero as the number of items increases, just as in the above section. Since this analysis offers no new insights from the previous one, it will be omitted.

\subsection{Multiplicative Bound}
The following section will prove the multiplicative bound for the asymptotic competitive ratio result. In what follows, all five properties above will be assumed, and the additive constants addressed above will be ignored. Thus, this section will provide an input to the dual linear program which produces an objective over $\frac{11}{16}$ times the number of bins used in complete sets of matched bins, complete sets of rematched bins, and additional excess Large, Medium, and Third Item bins.

\paragraph{Case 1 - No Adjustment}
Suppose during the algorithms execution, every Large Item is either rematched with Medium/Third Items or rematched into a Small Item complete set. Further suppose that every complete set of Third Items is rematched in some manner, whether with Large Items, Large Items and Quarter Items, a complete set of Small Items, or Quarter Items and a complete set of Small Items.

Complete matched Quarter Item sets have an average fullness of at least 0.6875, as do any other matched complete sets. Further, every rematched complete set will have an average fullness of at least 0.6875, as will rematched sets containing Quarter Items. Medium Items matched with Large Items have a per bin fullness over 0.6875, as do bins containing two Medium Items. Finally, bins containing one Very Large Item are at least 0.6875 full. Thus, for each item $i$, one can set $x_i = w_i$ in the dual to get a dual objective of at least $\frac{11}{16}$ times the number of bins used.

\paragraph{Case 2 - Excess of Third Items}
Suppose during the algorithms execution, every Large Item is either rematched with Medium Items or rematched into a Small Item complete set. Further suppose that not every complete set of Third Items is rematched. This will imply every complete set of Small Items (including potentially Quarter Item sets by themselves) are in complete sets of rematched bins.

In this case, it is possible to use the dual program to achieve the necessary bound. Set $x_i = 1/2$ for all Medium Items and Large Items matched with Medium Items, and set $x_i = 1$ for unmatched Large Items and Very Large Items. Set $x_i = 0$ for the remaining items. This will be a valid input to the dual program by the balanced bins property above. By the fullness properties, every set will have 0.6875 Large Items or 1.375 Third Items per bin, so will contribute 0.6875 to the dual per bin. Every other bin will contain either 2 Medium Items, a Medium Item and Large Item matched, or 1 Large Item, and thus contribute 1 to the dual objective per bin. Therefore, every bin will on average contribute at least $0.6875$ to the dual objective, so this objective will be at least $\frac{11}{16}$ times the number of bins the algorithm uses.

\paragraph{Case 3 - Excess of Quarter Items}
Suppose there are complete Quarter Item sets directly rematched with Large Items or Third Items, but not every such set is rematched. This will imply that every Large and Third Item set is rematched with other items in this instance. Further, every non-quarter Small Item category will be in rematched complete sets of bins and every item size under $1/4$ will be rematched with Quarter Item sets.

In this case, it is possible to set $x_i = 0.3125$ for every item with size in $(1/4, 1/3]$, Third Items, and Medium Items matched with Large Items, $x_i = 0.34375$ for Medium Items not matched with Large Items, $x_i = 0.5$ for Large Items matched with a Medium Item, and $x_i = 0.6875$ for Large Items not matched with a Medium Item. The balanced bins property will ensure this is a valid input to the dual program.

With these dual variable values, the quarter size reallocation property will guarantee that every complete set of rematched bins contributes at least 0.6875 the the dual objective per bin. Further, every bin containing two Medium Items will contribute 0.6875 to the objective, and every bin containing a Medium and Large Item will contribute well over this amount. Therefore, every bin will on average contribute at least $0.6875$ to the dual objective, so this objective will be at least the $\frac{11}{16}$ times the number of bins the algorithm uses.

\paragraph{Case 4 - Unmatchable Large Items}
Suppose that every Small Item set is rematched with Large Items, but not all such sets are rematched with Quarter Items. Further, suppose there are Medium Items or Third Item sets that cannot be matched with Large Items, and not enough Large Items to assign 11 Large Items for every 5 bins of unmatched Third and Medium Items. 

Take the smallest unmatched Medium Item or smallest Third Item in a complete set not rematched with Large Items, and assume it has a weight of $1-w$. This implies every Large Item not matched with a Medium Item will have size at least $w$ by the balanced bins property. Now, for every 5 bins containing two Medium or Third Items, there will need to be 11 bins containing 1 Large Item for every set to have at least $0.6875$ Large Items per bin. Further, this set must contribute 0.6875 per bin to the dual value. Thus, each Large Item not paired with a Medium Item will have its dual variable set to some value $w+x$. Solving $(2(1-w) \cdot 5 + 11(w+x))/16 = 0.6875$, it must be that $x = 1/11 - w/11$.

Each Large Item not paired with a Medium Item will have $x_i = 10w_i/11 + 1/11$. Medium Items not paired with a Large Item will have $x_i = w_i$, as will Large Items paired with Medium Items. For the remaining items to fit, it must be that $x_i = \frac{1 - 10w/11 - 1/11}{1 - w} w_i = 10 w_i/11$. A bin containing a Medium and Large Item will contribute at least $1/2 + 0.34375 \cdot 10/11 > 0.6875$ to the dual objective. Further, a set of Third Items rematched with Large Items will have 6 Large Items matched with Third Items and 1 Large Item not, so will contribute at least $(0.3125 \cdot 8 \cdot 10/11 + 0.5 \cdot 4 + .5 \cdot 10/11 + 1/11)/7 > 0.6875$ per bin to the dual objective. For all other complete rematched sets, they will have an average fullness of at least 0.6875 by properties 2 and 3. Thus, for every 16 bins in complete rematched sets with Small Items, there will be at least 11 Large Items and a density of at least 0.6875 per bin, leading to at least $(10(0.6875 \cdot 16 - 11w)/11 + 11(10w/11 + 1/11))/16  = 0.6875$ being contributed to the dual objective per bin. Therefore, every bin will on average contribute at least $0.6875$ to the dual objective, so this objective will be at least $\frac{11}{16}$ times the number of bins the algorithm uses.

\paragraph{Case 5 - Excess Large Items}
In the last case, suppose there are enough Large Items to rematched with every other item category and group with unmatched Medium or Third Items with some Large Items left over. Then, simply setting $x_i = 1$ for each item of size above 0.5 and $x_i = 0$ will contribute over 0.6875 per bin to the dual objective by the rematched fullness property, the Quarter Item rematched fullness property, and the manner with which Large Items are grouped with unmatched Medium and Third Items. So this objective will be at least $\frac{11}{16}$ times the number of bins the algorithm uses.





\paragraph{Completing the proof of Theorem 1} The work in Section 3.5 is sufficient to show that for any online MPAS instance $I$, 
\[\frac{11}{16}(ALG(I) - K(I)) \leq D \]
with $K(I)$ being all of the bins ignored in Sections 3.3 and 3.4 and $ALG(I)$ being the number of bins the algorithm uses after its rematching process. Bins ignored in Section 3.3 are bounded above by a constant, and bins ignored in Section 3.4 will almost surely be a vanishing fraction of the optimal number of bins used. Thus, $\lim_{OPT(I) \rightarrow \infty} \frac{K(I)}{OPT(I)} = 0$, satisfying the condition set out in this section's introduction. Therefore, the asymptotic competitive ratio of the algorithm is at most $\frac{16}{11}$, proving the claim.

\section{Lower Bounds}

This section will prove a lower bound of 1.2 on the asymptotic competitive ratio for any deterministic online MPAS algorithms. Then, a similar technique will be used to prove no randomized algorithm can achieve an asymptotic competitive ratio below 1.2 in expectation.

\subsection{Deterministic Lower Bound}

The same family of inputs is used to construct both the deterministic and randomized lower bounds for the MPAS problem. Though the deterministic result is implied by the randomized result, a deterministic lower bound is easier to prove and conceptualize. Thus, this subsection will be dedicated to proving the theorem below, before the full result in the next subsection.

\begin{theorem} [Asymptotic Deterministic Lower Bound]
Any deterministic algorithm for the online MPAS problem must have an asymptotic competitive ratio of at least $\frac{6}{5} = 1.2$.
\end{theorem}

\proof

Consider the set of inputs to the MPAS problem restricted to items of size $\frac{1}{3}$ and $\frac{2}{3}$. For any such input, there must be an optimal solution that only places items of size $\frac{1}{3}$ on either edge or in the exact center and only places items of size $\frac{2}{3}$ on an edge. Moving an item of size $\frac{2}{3}$ from wherever it is placed to its closest edge can not increase the number of items scheduled at the same time, since each job is at least of size $\frac{1}{3}$. The same is true for items of size $\frac{1}{3}$ not placed exactly in the middle of a bin.

Now, take any deterministic algorithm for the online MPAS problem which only schedules $\frac{1}{3}$ items in the middle or on an edge, and $\frac{2}{3}$ sized items on an edge, and choose some integer $n$. Consider an input to the algorithm with $n$ items of size $\frac{1}{3}$. If the algorithm schedules at least $\frac{1}{5}$ of those items in the center, then subsequently send $n$ items of size $\frac{2}{3}$; otherwise the input consists only of the first $n$ items. 

If under $\frac{1}{5}$ of the $\frac{1}{3}$ sized items were placed in the middle, then at least $\frac{2}{5}$ of the items would have been placed on either side. Thus, the algorithm will have at least $\frac{2n}{5}$ items at the peak appointment time, whereas the off-line optimum would have packed 3 items to a bin for a peak appointment of $\frac{n}{3}$. Thus, the algorithm must have achieved a competitive ratio of at least $\frac{6}{5}$ on this input.

Conversely, if over $\frac{1}{5}$ of the $\frac{1}{3}$-sized items were placed in the middle, then there must be at least $n + \frac{n}{5}$ items scheduled at the middle point in the day. But, an optimal solution would schedule each $\frac{2}{3}$ item with a $\frac{1}{3}$ item, for a peak appointment of $n$. Thus, the algorithm would have achieved a competitive ratio of at least $\frac{6}{5}$ on this input.

Since these properties hold for all $n$, taking $n$ to infinity and using this same pattern will create a sequence of inputs with an optimal solution tending toward infinity where every algorithm achieves a competitive ratio of at least $\frac{6}{5}$ on every input. Thus, the asymptotic competitive ratio of any deterministic algorithm must be at least $\frac{6}{5}$.

\subsection{Randomized Lower Bound}

A lower bound on the performance of any randomized algorithm for the MPAS problem will be developed in this section by first analyzing a simpler request-answer game (see, e.g., \cite{borodin2005online}). Consider a game where a requester can request an item be placed, either of size $\frac{1}{3}$ or $\frac{2}{3}$, and a responder needs to answer with a location within a bin to place the item. Further, define the cost function for this game to be the peak utilization within the bin, as in the MPAS problem. It is easy to see that this game is equivalent to the MPAS problem with item sizes restricted to $\frac{1}{3}$ and $\frac{2}{3}$, and thus any lower bounds on the game's asymptotic competitive ratio will bound the asymptotic competitive ratio of the MPAS problem.

\begin{theorem}[Request-Answer Asymptotic Randomized Lower Bound]
The above request-answer game has an asymptotic competitive ratio of at least $\frac{6}{5} = 1.2$.
\end{theorem}

\proof

Let $\sigma_n$ be the family of request sequences consisting of $n$ consecutive $\frac{1}{3}$ requests, and let $\gamma_n$ be the family of request sequences consisting of $n$ consecutive $\frac{1}{3}$ requests followed by $n$ consecutive $\frac{2}{3}$ requests. Let $y_n$ be a family of probability distributions over request sequences, with $y_n(\sigma_n) = \frac{2}{5}$, $y_n(\gamma_n) = \frac{3}{5}$, and $y_n(j) = 0$ otherwise. 

Choose some $n$ a multiple of 3. Now, take any deterministic algorithm $ALG$ for the responder, which when presented with $n$ consecutive items of size $\frac{1}{3}$, places $c$ of them in the middle position of a bin. Since every bin with 3 items must have an item in the middle and every item in the middle of a bin will have to conflict with an item of size $\frac{2}{3}$, the following must be true:

\[E_{y \in y_n}\left[\frac{ALG(y)}{OPT(y)} \right] = \frac{2}{5}\frac{ALG(\sigma_n)}{OPT(\sigma_n)} + \frac{3}{5}\frac{ALG(\gamma_n)}{OPT(\gamma_n)} \geq \frac{2}{5}\frac{c + (n-3c)/2}{n/3} + \frac{3}{5}\frac{n + c}{n} = \frac{6}{5}\]

By Yao's principle, it must be that the competitive ratio of any randomized algorithm for this game is at most $\frac{6}{5}$ against an oblivious adversary \cite{borodin2005online}. Further, the above inequality will hold for any $n$ which is a multiple of 3, which allows for the creation of arbitrarily large hard inputs. Adding a lower bound to the game on the cost of valid inputs will not change the lower bound, since one can just take $n$ to be sufficiently large to satisfy the cost lower bound. Thus, the competitive ratio must remain bounded below by $\frac{6}{5}$ for any minimum optimal cost, implying the asymptotic competitive ratio for this problem is at least $\frac{6}{5}$ as well.

This result will directly imply the main theorem for this section, by the line of reasoning at the start of this section.

\begin{theorem} [Asymptotic Randomized Lower Bound]
Any random algorithm for the online MPAS problem must have an asymptotic competitive ratio of at least $\frac{6}{5} = 1.2$.
\end{theorem}

\section{Conclusions and Future Work}

The work done in this paper provides an improvement to the asymptotic competitive ratio of the MPAS problem compared to previous work. It further proves that 1.5 is not a lower bound for this problem, and invites even further improvements. Moving forward, proving a lower bound on the asymptotic competitive ratio would help future analysis of this problem be more focused and guided. It is also a hope to reimagine other known online problems and study how decisions can be delayed and what properties emerge from such analysis.
\medskip\\

\bibliographystyle{abbrv}
\bibliography{template}  






\newpage
\appendix

\section{Particularly Hard Input}

It is possible to show that the asymptotic competitive ratio derived for the given algorithm is nearly the best possible. For any integer $n$, construct an input to the MPAS problem containing $2n$ items of size $0.3438$, $n$ items of size $0.2501$, and $n$ items of size $0.0623$ for some small epsilon. Since this input has no Large or Third Items, the algorithm will pack this input in the same manner regardless of the order items arrive. The optimal way to pack these items is to put two of the items of size $0.3438$ in a bin with one of each of the other items, requiring exactly $n$ bins. However, as $n$ goes to infinity, the algorithm will pack sets of two items of size $0.3438$ to a bin, sets of 11 items of size $0.2501$ to 4 bins, and sets of 54 items of size 0.0623 to 4 bins. This will result in the algorithm using roughly $n(1 + 4/11 + 4/54) = n\frac{427}{297}$ bins, for a competitive ratio of $\frac{427}{297} > 1.437$.

\section{Configuration Strategies}
 This section will detail a few important configuration strategies, which are used implicitly in the following section.
 
 \subsection{Alternating Sides}
 
Every complete set will have an even number of type 2 bins, and in the following section half of them will be on either side. Further, when rematching with Large Items, half of the Large Items will be on each side.
 
 \subsection{Type 2 Small Bins}
 
 When rematching a Small Item category with Large or Third Items, any type 2 small bins within the complete set of bins will be placed in a bin with a Large Item or inner bin respectively. 
 
 \subsection{Rematching with Different Large Item Categories}
 
 Several of the Small Item categories will have different rematching guidelines depending on the size of the Large Items (when rematching with only Large Items). In every such case, the smaller size will have every type 2 large bins are assigned to the same bin as a Large Item, while the larger case will have type 2 large bins assigned to a bin with another type 2 large bin.
 
 \subsection{Rematching with Quarter Item Sets}
 
 When rematching Small Item sets with a Quarter Item set and some other items, any type 2 large bins in the Small Item set will be rematched with the type 2 large Quarter Item bins if possible. This will leave the Quarter Item type 2 small bins to be rematched with Large Items or inner bins.
 
 \subsection{Rematching with Large Items}
 
 The properties the above analysis uses relies on there being 0.6875 Large Items for every bin after a rematching with a Small Item category, with an average fullness of at least 0.6875. In what follows, what will be shown is how to rematch so there are at least two Large Items per 3 bins, and an average density of at least 0.7. This is an equivalent statement, since for every 15 rematched bins an extra Large Item in its own bin can be assigned, leading to 11 Large Items for 16 bins with an average fullness of at least 11/16.

\section{Category List and Matching Process Description}
This section will be dedicated to going through each Small Item category, detailing the manner in which they are packed, and verifying all relevant properties used above for the category.

\subsection{Very Large Items}

Very Large Items will only have one category, and do not get rematched.

\subsection{Large Items}

The interior cutoffs for the Large Item categories will be: \\
$\frac{2}{3}, 0.642, 0.635, 0.625, 0.6, 0.588, 0.57, \frac{6}{11}, 0.54$

Rematching with Large Items is detailed in the relevant item category.

\subsection{Medium Items}

The interior cutoffs for the Medium Item categories will be: \\
$0.358, 0.365, 0.375, 0.4, 0.412, 0.43, \frac{5}{11}, 0.46$

Medium Items rematch with Large Items by being placed in the same bin as a Large Item on the opposite side, if they both fit in a bin.

Two Non-half Medium Items can rematch with a Quarter Item set and 8 Large Items. The two Medium Items are placed in a bin with the two Quarter Item type 2 large bins. The Quarter Item type 1 bin is placed in its own bin. The 4 Quarter Item type 2 small bins are placed in bins with Large Items, and the remaining 4 Large Items are placed in their own bin. This process will require 20 bins, and lead to an average fullness of at least $(0.34375 \cdot 2 + 0.25 \cdot 11 + 0.54 \cdot 8)/11 > 0.7$ over the bins. This process requires the Large Items to not fit in a bin with the relevant Medium Items remaining (otherwise this rematching would not happen in the algorithm's execution).

\subsection{Third Items}

Third Items will only have one category. 

Third Item sets can be rematched with 6 Large Items. Four left and right Large Items will be placed in the same bins as right and left outer bins respectively. The remaining 2 Large Items will be placed alone in a bin, as will the two inner bins. This process will require 8 bins, and lead to an average fullness of at least $((1/3)\cdot 8 + .5 \cdot 6)/8 > 0.7$ over the bins. This process requires there to be Large Items which fit in a bin with the relevant Third Items remaining.

Third Item sets can be rematched with 1 Quarter Item set and 14 Large Items. Two left and right Quarter Item type 2 large bins will be placed in the same bins as right and left outer bins respectively. The 4 type 2 small Quarter Item bins will be placed in a bin with a Large Item. The Quarter Item type 1 bins will remain in their own bin, as will the inner bins. The remaining two outer bins will be placed in their own bin. The remaining 10 Large Items will be placed alone in a bin. This process will require 20 bins, and lead to an average fullness of at least $((1/3)\cdot 8 + 0.25 \cdot 11 + (2/3) \cdot 14)/20 > 0.7$ over the bins. This process requires the Large Items to not fit in a bin with the relevant Third Items remaining (otherwise this rematching would not happen in the algorithm's execution).

\subsection{Small Items}

\subsubsection{Sup-Category 3 (0.3125, 1/3]}
A \textbf{Type 2 Large Bin} for this category will contain 1 item.\\
A \textbf{Type 1 Bin} for this category will contain 3 items.

A \textbf{Complete Set of Matched Bins} for this category will contain \\
- 2 Type 2 Large Bins Matched Pairwise \\
- 2 Type 1 Bins Unmatched\\
Filling 3 bins for an average fullness of at least $(0.3125 \cdot 8)/3 \geq 0.6875$

A \textbf{Complete Set of Rematched Bins} can be obtained by rematching with \\
4 Large Items under size 2/3 in 6 bins, for an average fullness of at least $(0.5 \cdot 4 + 0.3125 \cdot 8)/6 \geq 0.7$\\
6 Large Items over size 2/3 in 9 bins, for an average fullness of at least $(\frac{2}{3} \cdot 6 + 0.3125 \cdot 8)/9 \geq 0.7$\\
2 Third Item sets (16 Third Items) in 11 bins, for an average fullness of at least $(\frac{1}{3} \cdot 16 + 0.3125 \cdot 8)/11 \geq 0.6875$\\

This category will rematch with 0 Quarter Items sets and satisfy the quarter size reallocation property.

\subsubsection{Category 3 (0.27, 0.3125]}
A \textbf{Type 2 Small Bin} for this category will contain 1 item.\\
A \textbf{Type 1 Bin} for this category will contain 3 items.

A \textbf{Complete Set of Matched Bins} for this category will contain \\
- 4 Type 2 Small Bins Matched Pairwise \\
- 3 Type 1 Bins Unmatched\\
Filling 5 bins for an average fullness of at least $(0.27 \cdot 13)/5 \geq 0.6875$

A \textbf{Complete Set of Rematched Bins} can be obtained by rematching with \\
6 Large Items in 9 bins, for an average fullness of at least $(0.5 \cdot 6 + 0.27 \cdot 13)/9 \geq 0.7$\\
2 Third Item sets (16 Third Items) in 11 bins, for an average fullness of at least $(\frac{1}{3} \cdot 16 + 0.27 \cdot 13)/11 \geq 0.6875$\\

This category will rematch with 0 Quarter Items sets and satisfy the quarter size reallocation property.

\subsubsection{Quarter Items (0.25, 0.27]}
A \textbf{Type 2 Small Bin} for this category will contain 1 item.\\
A \textbf{Type 2 Large Bin} for this category will contain 2 items.\\
A \textbf{Type 1 Bin} for this category will contain 3 items.

A \textbf{Complete Set of Matched Bins} for this category will contain \\
- 2 Type 2 Small Bins Matched Pairwise \\
- 2 Type 2 Small Bins Matched with 2 Type 2 Large Bins Pairwise \\
- 1 Type 1 Bin Unmatched\\
Filling 4 bins for an average fullness of at least $(0.25 \cdot 11)/4 = 0.6875$

This category rematches in a unique way with almost every other item type, explained where relevant. When matched directly with Large Items, Quarter Item and Large Item bins are grouped together, with no items being rematched into new bins.

\subsubsection{Sup-Category 4 (0.23, 0.25]}
A \textbf{Type 2 Small Bin} for this category will contain 1 item.\\
A \textbf{Type 1 Bin} for this category will contain 4 items.

A \textbf{Complete Set of Matched Bins} for this category will contain \\
- 4 Type 2 Small Bins Matched Pairwise \\
- 2 Type 1 Bins Unmatched\\
Filling 4 bins for an average fullness of at least $(0.23 \cdot 12)/4 \geq 0.6875$

A \textbf{Complete Set of Rematched Bins} can be obtained by rematching with \\
4 Large Items in 6 bins, for an average fullness of at least $(0.5 \cdot 4 + 0.23 \cdot 12)/6 \geq 0.7$\\
2 Third Item sets (16 Third Items) in 10 bins, for an average fullness of at least $(\frac{1}{3} \cdot 16 + 0.23 \cdot 12)/10 \geq 0.6875$\\
10 Large Items \& 1 Quarter Item set in 15 bins, for an average fullness of at least $(0.5 \cdot 10 + 0.25 \cdot 11 + 0.23 \cdot 12)/15 \geq 0.7$\\ \hspace*{18 pt} and satisfying $(0.3125 \cdot 11 + 0.6875 \cdot 10)/15 \geq 0.6875$  \\
3 Third Item sets (24 Third Items) \& 1 Quarter Item set in 17 bins, for an average fullness of at least\\ \hspace*{18 pt} $(\frac{1}{3} \cdot 24 + 0.25 \cdot 11 + 0.23 \cdot 12)/17 \geq 0.6875$ and satisfying $(0.3125 \cdot 11 + 0.34375 \cdot 24)/17 \geq 0.6875$ 

\subsubsection{Category 4 (0.215, 0.23]}
A \textbf{Type 2 Small Bin} for this category will contain 1 item.\\
A \textbf{Type 2 Large Bin} for this category will contain 2 items.

A \textbf{Complete Set of Matched Bins} for this category will contain \\
- 2 Type 2 Small Bins Matched Pairwise \\
- 4 Type 2 Large Bins Matched Pairwise\\
Filling 3 bins for an average fullness of at least $(0.215 \cdot 10)/3 \geq 0.6875$

A \textbf{Complete Set of Rematched Bins} can be obtained by rematching with \\
6 Large Items under size 0.54 in 6 bins, for an average fullness of at least $(0.5 \cdot 6 + 0.215 \cdot 10)/6 \geq 0.7$\\
4 Large Items over size 0.54 in 6 bins, for an average fullness of at least $(0.54 \cdot 4 + 0.215 \cdot 10)/6 \geq 0.7$\\
2 Third Item sets (16 Third Items) in 10 bins, for an average fullness of at least $(\frac{1}{3} \cdot 16 + 0.215 \cdot 10)/10 \geq 0.6875$\\
8 Large Items \& 1 Quarter Item set in 12 bins, for an average fullness of at least $(0.5 \cdot 8 + 0.25 \cdot 11 + 0.215 \cdot 10)/12 \geq 0.7$\\ \hspace*{18 pt} and satisfying $(0.3125 \cdot 11 + 0.6875 \cdot 8)/12 \geq 0.6875$  \\
3 Third Item sets (24 Third Items) \& 1 Quarter Item set in 16 bins, for an average fullness of at least\\ \hspace*{18 pt} $(\frac{1}{3} \cdot 24 + 0.25 \cdot 11 + 0.215 \cdot 10)/16 \geq 0.6875$ and satisfying $(0.3125 \cdot 11 + 0.34375 \cdot 24)/16 \geq 0.6875$ 

\subsubsection{Sub-Category 4 (0.206, 0.215]}
A \textbf{Type 2 Small Bin} for this category will contain 1 item.\\
A \textbf{Type 2 Large Bin} for this category will contain 2 items.

A \textbf{Complete Set of Matched Bins} for this category will contain \\
- 2 Type 2 Small Bins Matched Pairwise \\
- 6 Type 2 Large Bins Matched Pairwise\\
Filling 3 bins for an average fullness of at least $(0.206 \cdot 14)/4 \geq 0.6875$

A \textbf{Complete Set of Rematched Bins} can be obtained by rematching with \\
8 Large Items under size 0.57 in 8 bins, for an average fullness of at least $(0.5 \cdot 8 + 0.206 \cdot 14)/8 \geq 0.7$\\
6 Large Items over size 0.57 in 9 bins, for an average fullness of at least $(0.57 \cdot 6 + 0.206 \cdot 14)/9 \geq 0.7$\\
2 Third Item sets (16 Third Items) in 11 bins, for an average fullness of at least $(\frac{1}{3} \cdot 16 + 0.206 \cdot 14)/11 \geq 0.6875$\\
14 Large Items \& 2 Quarter Item sets in 21 bins, for an average fullness of at least $(0.5 \cdot 14 + 0.25 \cdot 22 + 0.206 \cdot 14)/21 \geq 0.7$\\ \hspace*{18 pt} and satisfying $(0.3125 \cdot 22 + 0.6875 \cdot 14)/21 \geq 0.6875$  \\
3 Third Item sets (24 Third Items) \& 1 Quarter Item set in 17 bins, for an average fullness of at least\\ \hspace*{18 pt} $(\frac{1}{3} \cdot 24 + 0.25 \cdot 11 + 0.206 \cdot 14)/17 \geq 0.6875$ and satisfying $(0.3125 \cdot 11 + 0.34375 \cdot 24)/17 \geq 0.6875$ 

\subsubsection{Sub-Sub-Category 4 (0.2, 0.206]}
A \textbf{Type 2 Small Bin} for this category will contain 1 item.\\
A \textbf{Type 2 Large Bin} for this category will contain 2 items.

A \textbf{Complete Set of Matched Bins} for this category will contain \\
- 2 Type 2 Small Bins Matched Pairwise \\
- 6 Type 2 Large Bins Matched Pairwise\\
Filling 3 bins for an average fullness of at least $(0.2 \cdot 14)/4 \geq 0.6875$

A \textbf{Complete Set of Rematched Bins} can be obtained by rematching with \\
8 Large Items under size 0.588 in 8 bins, for an average fullness of at least $(0.5 \cdot 8 + 0.2 \cdot 14)/8 \geq 0.7$\\
6 Large Items over size 0.588 in 9 bins, for an average fullness of at least $(0.588 \cdot 6 + 0.2 \cdot 14)/9 \geq 0.7$\\
2 Third Item sets (16 Third Items) in 11 bins, for an average fullness of at least $(\frac{1}{3} \cdot 16 + 0.2 \cdot 14)/11 \geq 0.6875$\\
14 Large Items \& 2 Quarter Item sets in 21 bins, for an average fullness of at least $(0.5 \cdot 14 + 0.25 \cdot 22 + 0.2 \cdot 14)/21 \geq 0.7$\\ \hspace*{18 pt} and satisfying $(0.3125 \cdot 22 + 0.6875 \cdot 14)/21 \geq 0.6875$  \\
3 Third Item sets (24 Third Items) \& 1 Quarter Item set in 17 bins, for an average fullness of at least\\ \hspace*{18 pt} $(\frac{1}{3} \cdot 24 + 0.25 \cdot 11 + 0.2 \cdot 14)/17 \geq 0.6875$ and satisfying $(0.3125 \cdot 11 + 0.34375 \cdot 24)/17 \geq 0.6875$

\subsubsection{Sup-Category 5 (0.1825, 0.2]}
A \textbf{Type 2 Small Bin} for this category will contain 1 item.\\
A \textbf{Type 2 Large Bin} for this category will contain 2 items.\\
A \textbf{Type 1 Bin} for this category will contain 5 items.

A \textbf{Complete Set of Matched Bins} for this category will contain \\
- 2 Type 2 Small Bins Matched Pairwise\\
- 4 Type 2 Large Bins Matched Pairwise\\
- 2 Type 1 Bins Unmatched\\
Filling 5 bins for an average fullness of at least $(0.1825 \cdot 20)/5 \geq 0.6875$

A \textbf{Complete Set of Rematched Bins} can be obtained by rematching with \\
6 Large Items under size 0.6 in 8 bins, for an average fullness of at least $(0.5 \cdot 6 + 0.1825 \cdot 20)/8 \geq 0.7$\\
8 Large Items over size 0.6 in 12 bins, for an average fullness of at least $(0.6 \cdot 8 + 0.1825 \cdot 20)/12 \geq 0.7$\\
3 Third Item sets (24 Third Items) in 16 bins, for an average fullness of at least $(\frac{1}{3} \cdot 24 + 0.1825 \cdot 20)/16 \geq 0.6875$\\
16 Large Items \& 2 Quarter Item sets in 24 bins, for an average fullness of at least $(0.5 \cdot 16 + 0.25 \cdot 22 + 0.1825 \cdot 20)/24 \geq 0.7$\\ \hspace*{18 pt} and satisfying $(0.3125 \cdot 22 + 0.6875 \cdot 16)/24 \geq 0.6875$  \\
5 Third Item sets (40 Third Items) \& 2 Quarter Item sets in 28 bins, for an average fullness of at least\\ \hspace*{18 pt} $(\frac{1}{3} \cdot 40 + 0.25 \cdot 22 + 0.1825 \cdot 20)/28 \geq 0.6875$ and satisfying $(0.3125 \cdot 22 + 0.34375 \cdot 40)/28 \geq 0.6875$

\subsubsection{Category 5 (0.179, 0.1825]}
A \textbf{Type 2 Small Bin} for this category will contain 1 item.\\
A \textbf{Type 2 Large Bin} for this category will contain 2 items.\\
A \textbf{Type 1 Bin} for this category will contain 5 items.

A \textbf{Complete Set of Matched Bins} for this category will contain \\
- 4 Type 2 Small Bins Matched Pairwise\\
- 10 Type 2 Large Bins Matched Pairwise\\
- 6 Type 1 Bins Unmatched\\
Filling 13 bins for an average fullness of at least $(0.179 \cdot 54)/13 \geq 0.6875$

A \textbf{Complete Set of Rematched Bins} can be obtained by rematching with \\
14 Large Items under size 0.635 in 20 bins, for an average fullness of at least $(0.5 \cdot 14 + 0.179 \cdot 54)/20 \geq 0.7$\\
22 Large Items over size 0.635 in 33 bins, for an average fullness of at least $(0.635 \cdot 22 + 0.179 \cdot 54)/33 \geq 0.7$\\
7 Third Item sets (56 Third Items) in 39 bins, for an average fullness of at least $(\frac{1}{3} \cdot 56 + 0.179 \cdot 54)/39 \geq 0.6875$\\
42 Large Items \& 5 Quarter Item sets in 63 bins, for an average fullness of at least $(0.5 \cdot 42 + 0.25 \cdot 55 + 0.179 \cdot 54)/63 \geq 0.7$\\ \hspace*{18 pt} and satisfying $(0.3125 \cdot 55 + 0.6875 \cdot 42)/63 \geq 0.6875$  \\
12 Third Item sets (96 Third Items) \& 5 Quarter Item sets in 69 bins, for an average fullness of at least\\ \hspace*{18 pt} $(\frac{1}{3} \cdot 96 + 0.25 \cdot 55 + 0.179 \cdot 54)/69 \geq 0.6875$ and satisfying $(0.3125 \cdot 55 + 0.34375 \cdot 96)/69 \geq 0.6875$

\subsubsection{Sub-Category 5 (1/6, 0.179]}
A \textbf{Type 2 Small Bin} for this category will contain 1 item.\\
A \textbf{Type 2 Large Bin} for this category will contain 2 items.\\
A \textbf{Type 1 Bin} for this category will contain 5 items.

A \textbf{Complete Set of Matched Bins} for this category will contain \\
- 4 Type 2 Small Bins Matched Pairwise\\
- 10 Type 2 Large Bins Matched Pairwise\\
- 6 Type 1 Bins Unmatched\\
Filling 13 bins for an average fullness of at least $((1/6) \cdot 54)/13 \geq 0.6875$

A \textbf{Complete Set of Rematched Bins} can be obtained by rematching with \\
14 Large Items under size 0.642 in 20 bins, for an average fullness of at least $(0.5 \cdot 14 + (1/6) \cdot 54)/20 \geq 0.7$\\
22 Large Items over size 0.642 in 33 bins, for an average fullness of at least $(0.642 \cdot 22 + (1/6) \cdot 54)/33 \geq 0.7$\\
7 Third Item sets (56 Third Items) in 39 bins, for an average fullness of at least $(\frac{1}{3} \cdot 56 + (1/6) \cdot 54)/39 \geq 0.6875$\\
36 Large Items under size 0.642 \& 4 Quarter Item sets in 54 bins, for an average fullness of at least\\ \hspace*{18 pt} $(0.5 \cdot 36 + 0.25 \cdot 44 + (1/6) \cdot 54)/54 \geq 0.7$ and satisfying $(0.3125 \cdot 44 + 0.6875 \cdot 36)/54 \geq 0.6875$  \\
42 Large Items over size 0.642 \& 5 Quarter Item sets in 63 bins, for an average fullness of at least\\ \hspace*{18 pt} $(0.642 \cdot 42 + 0.25 \cdot 55 + (1/6) \cdot 54)/63 \geq 0.7$ and satisfying $(0.3125 \cdot 55 + 0.6875 \cdot 42)/63 \geq 0.6875$  \\
12 Third Item sets (96 Third Items) \& 5 Quarter Item sets in 69 bins, for an average fullness of at least\\ \hspace*{18 pt} $(\frac{1}{3} \cdot 96 + 0.25 \cdot 55 + (1/6) \cdot 54)/69 \geq 0.6875$ and satisfying $(0.3125 \cdot 55 + 0.34375 \cdot 96)/69 \geq 0.6875$

\subsubsection{Sup-Category 6 (0.15625, 1/6]}
A \textbf{Type 2 Small Bin} for this category will contain 1 item.\\
A \textbf{Type 2 Large Bin} for this category will contain 2 items.\\
A \textbf{Type 1 Bin} for this category will contain 6 items.

A \textbf{Complete Set of Matched Bins} for this category will contain \\
- 2 Type 2 Small Bins Matched Pairwise \\
- 2 Type 2 Large Bins Matched Pairwise\\
- 2 Type 1 Bins Unmatched\\
Filling 4 bins for an average fullness of at least $(0.15625 \cdot 18)/4 \geq 0.6875$

A \textbf{Complete Set of Rematched Bins} can be obtained by rematching with \\
4 Large Items under size 2/3 in 6 bins, for an average fullness of at least $(0.5 \cdot 4 + 0.15625 \cdot 18)/6 \geq 0.7$\\
6 Large Items over size 2/3 in 9 bins, for an average fullness of at least $((2/3) \cdot 6 + 0.15625 \cdot 18)/9 \geq 0.7$\\
2 Third Item sets (16 Third Items) in 11 bins, for an average fullness of at least $(\frac{1}{3} \cdot 16 + 0.15625 \cdot 18)/11 \geq 0.6875$\\
10 Large Items \& 1 Quarter Item set in 15 bins, for an average fullness of at least $(0.5 \cdot 10 + 0.25 \cdot 11 + 0.15625 \cdot 18)/15 \geq 0.7$\\ \hspace*{18 pt} and satisfying $(0.3125 \cdot 11 + 0.6875 \cdot 10)/15 \geq 0.6875$  \\
3 Third Item sets (24 Third Items) \& 1 Quarter Item set in 17 bins, for an average fullness of at least\\ \hspace*{18 pt} $(\frac{1}{3} \cdot 24 + 0.25 \cdot 11 + 0.15625 \cdot 18)/17 \geq 0.6875$ and satisfying $(0.3125 \cdot 11 + 0.34375 \cdot 24)/17 \geq 0.6875$ 

\subsubsection{Category 6 (1/7, 0.15625]}
A \textbf{Type 2 Small Bin} for this category will contain 2 items.\\
A \textbf{Type 1 Bin} for this category will contain 6 items.

A \textbf{Complete Set of Matched Bins} for this category will contain \\
- 4 Type 2 Small Bins Matched Pairwise \\
- 2 Type 1 Bins Unmatched\\
Filling 4 bins for an average fullness of at least $((1/7) \cdot 20)/4 \geq 0.6875$

A \textbf{Complete Set of Rematched Bins} can be obtained by rematching with \\
4 Large Items in 6 bins, for an average fullness of at least $(0.5 \cdot 4 + (1/7) \cdot 20)/6 \geq 0.7$\\
2 Third Item sets (16 Third Items) in 10 bins, for an average fullness of at least $(\frac{1}{3} \cdot 16 + (1/7) \cdot 20)/10 \geq 0.6875$\\
10 Large Items \& 1 Quarter Item set in 15 bins, for an average fullness of at least $(0.5 \cdot 10 + 0.25 \cdot 11 + (1/7) \cdot 20)/15 \geq 0.7$\\ \hspace*{18 pt} and satisfying $(0.3125 \cdot 11 + 0.6875 \cdot 10)/15 \geq 0.6875$  \\
3 Third Item sets (24 Third Items) \& 1 Quarter Item set in 17 bins, for an average fullness of at least\\ \hspace*{18 pt} $(\frac{1}{3} \cdot 24 + 0.25 \cdot 11 + (1/7) \cdot 20)/17 \geq 0.6875$ and satisfying $(0.3125 \cdot 11 + 0.34375 \cdot 24)/17 \geq 0.6875$ 

\subsubsection{Category 7 (1/8, 1/7]}
A \textbf{Type 2 Small Bin} for this category will contain 2 item.\\
A \textbf{Type 2 Large Bin} for this category will contain 3 items.\\
A \textbf{Type 1 Bin} for this category will contain 7 items.

A \textbf{Complete Set of Matched Bins} for this category will contain \\
- 4 Type 2 Small Bins Matched Pairwise \\
- 2 Type 2 Large Bins Matched Pairwise\\
- 2 Type 1 Bins Unmatched\\
Filling 5 bins for an average fullness of at least $((1/8) \cdot 28)/5 \geq 0.6875$

A \textbf{Complete Set of Rematched Bins} can be obtained by rematching with \\
6 Large Items in 9 bins, for an average fullness of at least $(0.5 \cdot 6 + (1/8) \cdot 28)/9 \geq 0.7$\\
2 Third Item sets (16 Third Items) in 11 bins, for an average fullness of at least $(\frac{1}{3} \cdot 16 + (1/8) \cdot 28)/11 \geq 0.6875$\\
10 Large Items \& 1 Quarter Item set in 15 bins, for an average fullness of at least $(0.5 \cdot 10 + 0.25 \cdot 11 + (1/8) \cdot 28)/15 \geq 0.7$\\ \hspace*{18 pt} and satisfying $(0.3125 \cdot 11 + 0.6875 \cdot 10)/15 \geq 0.6875$  \\
4 Third Item sets (32 Third Items) \& 1 Quarter Item set in 21 bins, for an average fullness of at least\\ \hspace*{18 pt} $(\frac{1}{3} \cdot 32 + 0.25 \cdot 11 + (1/8) \cdot 28)/21 \geq 0.6875$ and satisfying $(0.3125 \cdot 11 + 0.34375 \cdot 32)/21 \geq 0.6875$ 

\subsubsection{Category 8 (1/9, 1/8]}
A \textbf{Type 2 Small Bin} for this category will contain 2 item.\\
A \textbf{Type 2 Large Bin} for this category will contain 3 items.\\
A \textbf{Type 1 Bin} for this category will contain 8 items.

A \textbf{Complete Set of Matched Bins} for this category will contain \\
- 2 Type 2 Small Bins Matched Pairwise \\
- 2 Type 2 Large Bins Matched Pairwise\\
- 2 Type 1 Bins Unmatched\\
Filling 4 bins for an average fullness of at least $((1/9) \cdot 26)/4 \geq 0.6875$

A \textbf{Complete Set of Rematched Bins} can be obtained by rematching with \\
4 Large Items under size 5/8 in 6 bins, for an average fullness of at least $(0.5 \cdot 4 + (1/9) \cdot 26)/6 \geq 0.7$\\
6 Large Items over size 5/8 in 9 bins, for an average fullness of at least $((5/8) \cdot 6 + (1/9) \cdot 26)/9 \geq 0.7$\\
2 Third Item sets (16 Third Items) in 11 bins, for an average fullness of at least $(\frac{1}{3} \cdot 16 + (1/9) \cdot 26)/11 \geq 0.6875$\\
10 Large Items \& 1 Quarter Item set in 15 bins, for an average fullness of at least $(0.5 \cdot 10 + 0.25 \cdot 11 + (1/9) \cdot 26)/15 \geq 0.7$\\ \hspace*{18 pt} and satisfying $(0.3125 \cdot 11 + 0.6875 \cdot 10)/15 \geq 0.6875$  \\
4 Third Item sets (32 Third Items) \& 1 Quarter Item set in 21 bins, for an average fullness of at least\\ \hspace*{18 pt} $(\frac{1}{3} \cdot 32 + 0.25 \cdot 11 + (1/9) \cdot 26)/21 \geq 0.6875$ and satisfying $(0.3125 \cdot 11 + 0.34375 \cdot 32)/21 \geq 0.6875$

\subsubsection{Category 9 (1/10, 1/9]}
A \textbf{Type 2 Small Bin} for this category will contain 2 item.\\
A \textbf{Type 2 Large Bin} for this category will contain 3 items.\\
A \textbf{Type 1 Bin} for this category will contain 9 items.

A \textbf{Complete Set of Matched Bins} for this category will contain \\
- 2 Type 2 Small Bins Matched Pairwise \\
- 2 Type 2 Large Bins Matched Pairwise\\
- 2 Type 1 Bins Unmatched\\
Filling 4 bins for an average fullness of at least $((1/10) \cdot 28)/4 \geq 0.6875$

A \textbf{Complete Set of Rematched Bins} can be obtained by rematching with \\
4 Large Items under size 2/3 in 6 bins, for an average fullness of at least $(0.5 \cdot 4 + (1/10) \cdot 28)/6 \geq 0.7$\\
6 Large Items over size 2/3 in 9 bins, for an average fullness of at least $((2/3) \cdot 6 + (1/10) \cdot 28)/9 \geq 0.7$\\
2 Third Item sets (16 Third Items) in 11 bins, for an average fullness of at least $(\frac{1}{3} \cdot 16 + (1/10) \cdot 28)/11 \geq 0.6875$\\
10 Large Items \& 1 Quarter Item set in 15 bins, for an average fullness of at least $(0.5 \cdot 10 + 0.25 \cdot 11 + (1/10) \cdot 28)/15 \geq 0.7$\\ \hspace*{18 pt} and satisfying $(0.3125 \cdot 11 + 0.6875 \cdot 10)/15 \geq 0.6875$  \\
4 Third Item sets (32 Third Items) \& 1 Quarter Item set in 21 bins, for an average fullness of at least\\ \hspace*{18 pt} $(\frac{1}{3} \cdot 32 + 0.25 \cdot 11 + (1/10) \cdot 28)/21 \geq 0.6875$ and satisfying $(0.3125 \cdot 11 + 0.34375 \cdot 32)/21 \geq 0.6875$

\subsubsection{Category 10 (1/11, 1/10]}
A \textbf{Type 2 Small Bin} for this category will contain 3 item.\\
A \textbf{Type 2 Large Bin} for this category will contain 4 items.\\
A \textbf{Type 1 Bin} for this category will contain 10 items.

A \textbf{Complete Set of Matched Bins} for this category will contain \\
- 2 Type 2 Small Bins Matched Pairwise \\
- 2 Type 2 Large Bins Matched Pairwise\\
- 2 Type 1 Bins Unmatched\\
Filling 4 bins for an average fullness of at least $((1/11) \cdot 34)/4 \geq 0.6875$

A \textbf{Complete Set of Rematched Bins} can be obtained by rematching with \\
4 Large Items under size 0.6 in 6 bins, for an average fullness of at least $(0.5 \cdot 4 + (1/11) \cdot 34)/6 \geq 0.7$\\
6 Large Items over size 0.6 in 9 bins, for an average fullness of at least $(0.6 \cdot 6 + (1/11) \cdot 34)/9 \geq 0.7$\\
2 Third Item sets (16 Third Items) in 11 bins, for an average fullness of at least $(\frac{1}{3} \cdot 16 + (1/11) \cdot 34)/11 \geq 0.6875$\\
10 Large Items \& 1 Quarter Item set in 15 bins, for an average fullness of at least $(0.5 \cdot 10 + 0.25 \cdot 11 + (1/11) \cdot 34)/15 \geq 0.7$\\ \hspace*{18 pt} and satisfying $(0.3125 \cdot 11 + 0.6875 \cdot 10)/15 \geq 0.6875$  \\
4 Third Item sets (32 Third Items) \& 1 Quarter Item set in 21 bins, for an average fullness of at least\\ \hspace*{18 pt} $(\frac{1}{3} \cdot 32 + 0.25 \cdot 11 + (1/11) \cdot 34)/21 \geq 0.6875$ and satisfying $(0.3125 \cdot 11 + 0.34375 \cdot 32)/21 \geq 0.6875$

\subsubsection{Category 11 (1/12, 1/11]}
A \textbf{Type 2 Small Bin} for this category will contain 3 item.\\
A \textbf{Type 2 Large Bin} for this category will contain 5 items.\\
A \textbf{Type 1 Bin} for this category will contain 11 items.

A \textbf{Complete Set of Matched Bins} for this category will contain \\
- 2 Type 2 Small Bins Matched Pairwise \\
- 2 Type 2 Large Bins Matched Pairwise\\
- 2 Type 1 Bins Unmatched\\
Filling 4 bins for an average fullness of at least $((1/12) \cdot 38)/4 \geq 0.6875$

A \textbf{Complete Set of Rematched Bins} can be obtained by rematching with \\
4 Large Items under size 6/11 in 6 bins, for an average fullness of at least $(0.5 \cdot 4 + (1/12) \cdot 38)/6 \geq 0.7$\\
6 Large Items over size 6/11 in 9 bins, for an average fullness of at least $((6/11) \cdot 6 + (1/12) \cdot 38)/9 \geq 0.7$\\
2 Third Item sets (16 Third Items) in 11 bins, for an average fullness of at least $(\frac{1}{3} \cdot 16 + (1/12) \cdot 38)/11 \geq 0.6875$\\
10 Large Items \& 1 Quarter Item set in 15 bins, for an average fullness of at least $(0.5 \cdot 10 + 0.25 \cdot 11 + (1/12) \cdot 38)/15 \geq 0.7$\\ \hspace*{18 pt} and satisfying $(0.3125 \cdot 11 + 0.6875 \cdot 10)/15 \geq 0.6875$  \\
4 Third Item sets (32 Third Items) \& 1 Quarter Item set in 21 bins, for an average fullness of at least\\ \hspace*{18 pt} $(\frac{1}{3} \cdot 32 + 0.25 \cdot 11 + (1/12) \cdot 38)/21 \geq 0.6875$ and satisfying $(0.3125 \cdot 11 + 0.34375 \cdot 32)/21 \geq 0.6875$

\subsubsection{Category 12 [0, 1/12]}
This category will be slightly different, since it includes all items of size at most 1/12. So each bin type will not specify a number of items, but rather a cutoff a which point items will stop being added. For example, a Type 1 bin will be until 11/12 full, which means that it will not be "full" and able to accept more items until it is 11/12ths full, at which point it may not be able to accept another item without going over the limit and is thus at capacity.

A \textbf{Type 2 Small Bin} for this category will be filled until 1/4 full.\\
A \textbf{Type 2 Large Bin} for this category will be filled until 113/300 full.\\
A \textbf{Type 1 Bin} for this category will be filled until 11/12 full.

A \textbf{Complete Set of Matched Bins} for this category will contain \\
- 2 Type 2 Small Bins Matched Pairwise \\
- 2 Type 2 Large Bins Matched Pairwise\\
- 2 Type 1 Bins Unmatched\\
Filling 4 bins for an average fullness of at least $(((1/4) + (113/300) + (11/12)) \cdot 2)/4 \geq 0.6875$

A \textbf{Complete Set of Rematched Bins} can be obtained by rematching with \\
4 Large Items under size 0.54 in 6 bins, for an average fullness of at least $(0.5 \cdot 4 + ((1/4) + (113/300) + (11/12)) \cdot 2)/6 \geq 0.7$\\
6 Large Items over size 0.54 in 9 bins, for an average fullness of at least $(0.54 \cdot 6 + ((1/4) + (113/300) + (11/12)) \cdot 2)/9 \geq 0.7$\\
2 Third Item sets (16 Third Items) in 11 bins, for an average fullness of at least\\ \hspace*{18 pt} $(\frac{1}{3} \cdot 16 + ((1/4) + (113/300) + (11/12)) \cdot 2)/11 \geq 0.6875$\\
10 Large Items \& 1 Quarter Item set in 15 bins, for an average fullness of at least\\ \hspace*{18 pt} $(0.5 \cdot 10 + 0.25 \cdot 11 + ((1/4) + (113/300) + (11/12)) \cdot 2)/15 \geq 0.7$ and satisfying $(0.3125 \cdot 11 + 0.6875 \cdot 10)/15 \geq 0.6875$  \\
4 Third Item sets (32 Third Items) \& 1 Quarter Item set in 21 bins, for an average fullness of at least\\ \hspace*{18 pt} $(\frac{1}{3} \cdot 32 + 0.25 \cdot 11 + ((1/4) + (113/300) + (11/12)) \cdot 2)/21 \geq 0.6875$ and satisfying $(0.3125 \cdot 11 + 0.34375 \cdot 32)/21 \geq 0.6875$

\end{document}